\documentclass[twocolumn,superscriptaddress,showpacs,nofootinbib,preprintnumbers,secnumarabic,amssymb, nobibnotes, aps, prd]{revtex4-2}
\usepackage[utf8]{inputenc}
\usepackage{graphicx}
\usepackage{latexsym,amsmath,amssymb,amsthm,lmodern,float,url}
\usepackage{natbib}
\usepackage{color}
\usepackage[usenames,dvipsnames]{xcolor}
\usepackage{microtype}
\usepackage{import}
\usepackage{bbold}
\usepackage[plain]{fancyref}
\usepackage{varioref}
\usepackage{slashed}
\usepackage{multirow}
\usepackage{tikz}
\usepackage{scrextend}
\usepackage{braket}
\usetikzlibrary{quantikz2}
\usetikzlibrary{shapes}
\usetikzlibrary{positioning}
\usepackage[normalem]{ulem}
\usepackage{mathtools}

\newcommand{\fig}[1]{Fig.~\ref{fig:#1}}
\newcommand{\tab}[1]{Tab.~\ref{tab:#1}}
\newcommand{\eq}[1]{Eq.~(\ref{eq:#1})}
\newcommand{\seca}[1]{App.~\ref{app:#1}}
\newcommand{\sect}[1]{Sec.~\ref{sec:#1}}
\definecolor{seagreen}{rgb}{0.180392,0.545098,0.341176}

\newcommand{\bi}{\mathbb{BI}}
\newcommand{\bo}{\mathbb{BO}}
\newcommand{\btt}{\mathbb{BT}}
\newcommand{\vg}{\mathbb{V}}
\usepackage[colorlinks=true,backref=false, linktocpage=true,
citecolor=blue,urlcolor=blue,linkcolor=blue,pdfpagemode=UseOutlines]{hyperref}
\hypersetup{
  bookmarksnumbered=true,
  pdftitle = {},
  pdfsubject = {},
  pdfauthor = {},
  pdfkeywords = {}
}

\DeclareMathOperator{\tr}{Tr}

\begin{document}
\preprint{USTC-ICTS/PCFT-24-15, FERMILAB-PUB-24-0242-T}
\title{Block Encodings of Discrete Subgroups on Quantum Computer} 
\author{Henry Lamm}
\email{hlamm@fnal.gov}
\affiliation{Fermi National Accelerator Laboratory, Batavia,  Illinois, 60510, USA}
\author{Ying-Ying Li}
\email{yingyingli@ustc.edu.cn}
\affiliation{Interdisciplinary Center for Theoretical Study,
University of Science and Technology of China, Hefei, Anhui 230026, China}
\affiliation{Peng Huanwu Center for Fundamental Theory, Hefei, Anhui 230026, China}
\author{Jing Shu}
\email{jshu@pku.edu.cn}
\affiliation{School of Physics and State Key Laboratory of Nuclear Physics and Technology, 
	Peking University, Beijing 100871, China}
\affiliation{Center for High Energy Physics, Peking University, Beijing 100871, China}
\affiliation{Beijing Laser Acceleration Innovation Center, Huairou, Beijing, 101400, China}
\author{Yi-Lin Wang}
\email{wangyilin@mail.ustc.edu.cn}
\affiliation{Interdisciplinary Center for Theoretical Study,
University of Science and Technology of China, Hefei, Anhui 230026, China}
\affiliation{Peng Huanwu Center for Fundamental Theory, Hefei, Anhui 230026, China}
\author{Bin Xu}
\email{binxu@pku.edu.cn}
\affiliation{School of Physics and State Key Laboratory of Nuclear Physics and Technology, Peking University, Beijing 100871, China}

\date{\today}

\begin{abstract}
We introduce a block encoding method for mapping discrete subgroups to qubits on a quantum computer. This method is applicable to general discrete groups, including crystal-like subgroups such as $\mathbb{BI}$ of $SU(2)$ and $\mathbb{V}$ of $SU(3)$. We detail the construction of primitive gates -- the inversion gate, the group multiplication gate, the trace gate, and the group Fourier gate -- utilizing this encoding method for $\mathbb{BT}$ and for the first time $\mathbb{BI}$ group. We also provide resource estimations to extract the gluon viscosity. The inversion gates for $\mathbb{BT}$ and $\mathbb{BI}$ are benchmarked on the \texttt{Baiwang} quantum computer with estimated fidelities of $40^{+5}_{-4}\%$ and $4^{+5}_{-3}\%$ respectively. 
\end{abstract}

\maketitle
\section{Introduction}
Gauge symmetries play important roles in quantum field theories, with the $SU(3) \times SU(2) \times U(1)$ gauge symmetry being particularly important as it encapsulates the interactions in the well-established Standard Model for particle physics. Accurate predictions in the strongly-coupled regime of these interactions require substantial computational resources. Over the past few decades, Monte Carlo methods in lattice gauge theory (LGT) have flourished, benefiting from advances in supercomputing capabilities and algorithmic innovations. Nevertheless, challenges persist, particularly in scenarios involving dynamic processes such as transport coefficients of the quark-gluon plasma~\cite{Cohen:2021imf,Turro:2024pxu,Farrell:2024mgu}, parton physics in hadron collisions \cite{Lamm:2019uyc,Kreshchuk:2020dla,Echevarria:2020wct,Bauer:2021gup,Li:2021kcs,Zache:2023cfj}, and out-of-equilibrium evolution in the early universe~\cite{Polkovnikov_2008, Baum:2020vfl, PhysRevLett.108.080402,Czajka:2021yll} due to sign problems from the complex-valued nature of the Boltzmann sampling weight. Quantum computers offer a promising avenue to circumvent this challenge by enabling real-time simulations within the Hamiltonian formalism~\cite{Feynman:1981tf,Jordan:2017lea,Banuls:2019bmf,Bauer:2022hpo, DiMeglio:2023nsa}.

The Hilbert space of LGT is infinite-dimensional, requiring digitization methods to facilitate its mapping onto a finite quantum memory (see Sec VI.b of~\cite{Bauer:2022hpo} for different digitization methods). These include the loop-string-hadron (LSH) formulations \cite{Anishetty_2009, PhysRevD.90.114503, PhysRevD.101.114502} where the Hilbert space is built from gauge-invariant operators, digitizations of independent Wilson loops \cite{Zohar:2013zla, PhysRevD.102.094515, PhysRevD.102.114517, Yamamoto:2020eqi, bauer2021efficient, ciavarella2024quantum}, qubitization formulations of gauge theories \cite{PhysRevD.102.114514, PhysRevLett.126.172001, Alexandru_2022}, and the focus of this paper -the discrete subgroup approximation\cite{PhysRevD.91.054506, PhysRevD.100.034518}. Understanding and reducing the theoretical errors in the digitization is an area of active research~\cite{Shaw:2020udc,Tong:2021rfv,Ji:2020kjk,PhysRevD.106.114504, Bauer:2023jvw,Ciavarella:2023mfc,Hanada:2022pps}.

\begin{figure}[ht]
\centering
    \includegraphics[width=0.9\linewidth]{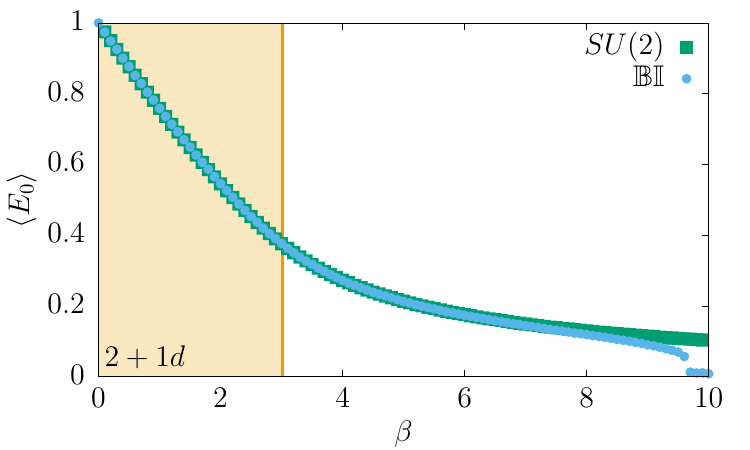}
    \includegraphics[width=0.9\linewidth]{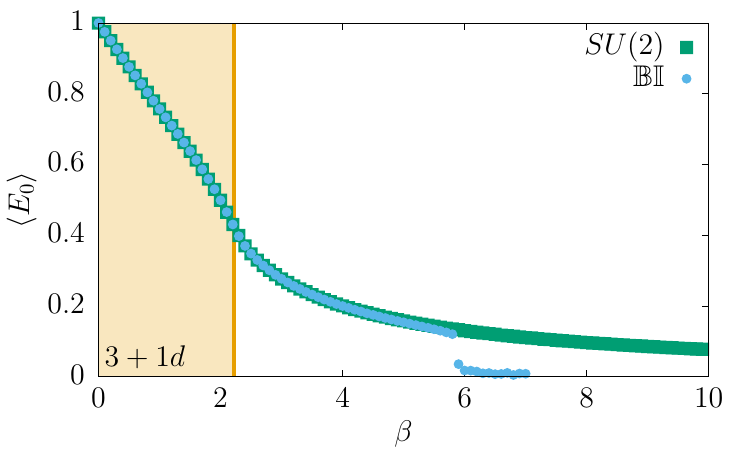}
    \caption{\label{fig:freezing} Average plaquette or lattice energy density $\langle E_0\rangle$ of $\bi$ as a function of Wilson coupling parameter $\beta$ on $8^d$ lattices for (top) $2+1d$ with $\beta^{2+1}_f=9.65(5)$ and (bottom) $3+1d$ with $\beta^{3+1}_{f}=5.85(5)$. The shaded region corresponds to the phase where $\beta\leq \beta_s$.} 
    \label{fig:lattice-energy}
\end{figure}

LGT calculations are performed at lattice spacing $a = a (\beta)$ which approaches zero for asymptomatic free theories as the coupling parameter $\beta\to \infty$. To perform extrapolation to the continuous spacetime limit, calculations need to be done in the scaling regime with $\beta > \beta_s$. For the discrete subgroup approximation, gauge links will become ``frozen'' to the identity when $\beta > \beta_f$, leading to different behaviors from the continuous groups, which makes the discrete subgroup approximations to be only valid in the regime $\beta < \beta_f$. Thus to extrapolate to the correct continuous spacetime limit as the continuous group, the discrete subgroup approximations need to satisfy the condition $\beta_f > \beta_s$. The discrete subgroup approximation has been explored for the Abelian group $U(1)$ \cite{Creutz:1979zg, Creutz:1982dn} and $SU(N)$ gauge theories \cite{Bhanot:1981xp, Petcher:1980cq, Bhanot:1981pj, Gustafson:2022xdt, PhysRevD.106.114504, Hackett:2018cel, Alam:2021uuq, Ji:2020kjk,Ji:2022qvr, Alexandru:2021jpm}, including fermions \cite{Weingarten:1980hx, Weingarten:1981jy}. For the case of $SU(2)$ which we consider in this work, there are three crystal-like subgroups: $\btt$, $\bo$, and $\bi$ with $\beta_f^d$ shown in \tab{groupcomp}. The smallest group, $\btt$ has $\beta_f^{3+1}<\beta_s^{3+1}\sim2.2$, while the other two subgroup $\bi$, $\beta^d_f \gg \beta^d_s$ as shown in \fig{lattice-energy} for both $2+1d$ and $3+1d$ theories. The larger $\beta_f$ of $\bi$ implies that digitization errors should be smaller in comparison to $\btt$ and $\bo$. Other attempts to increase $\beta_f$ with modified actions are also being discussed \cite{PhysRevD.100.114501}.

\begin{table}
\caption{\label{tab:groupcomp} Freezing couplings as a function of spatial dimension $d$, $\beta_f^{d+1}$, for crystal-like subgroups of $SU(2)$. For comparison, $\beta_s^{2+1d}=3$ and $\beta_s^{2+1d}=2.2$.}
\begin{center}
\begin{tabular}
{c| c c}
\hline\hline
$G$ & $\beta^{2+1d}_f$ & $\beta^{3+1d}_f$ \\
\hline
$\btt$\footnote{\label{refbt}Numerical results from \cite{Gustafson:2022xdt}} &3.45(5)&2.25(5)\\
$\bo$\footnote{\label{refbo}Numerical results from \cite{Gustafson:2023kvd}}&5.45(5)&3.25(5)\\
$\bi$&9.65(5)&5.85(5)\\
\hline\hline
\end{tabular}
\end{center}
\end{table}

Though reducing the uncertainties in the extrapolation to the continuum with larger $\beta_f$, quantum simulations of larger discrete group will be more complicated. To simulate the discrete subgroup theory on a quantum computer, one need to obtain the mapping between the elements of the discrete group and the qubits. The usual way is to write the elements of certain discrete subgroup as an ordered product of group generators with the exponents mapped to qubits or qudits on quantum computer \cite{Gustafson:2022xdt, Gustafson:2023kvd,Gustafson:2023swx,Gustafson:2024kym}.
However, for the largest crystal-like $\bi$ and $\vg$ subgroups of $SU(2)$ and $SU(3)$ theories, the ordered product expression is unknown yet, and may not be possible. This potential obstacle motivates the search for other encodings of discrete groups.

In this work, we consider a new encoding -- \emph{block encoding} -- expressing groups as $d$-dimensional matrices over a finite field $\mathbf{F}$. According to Cayley's theorem, any finite group of order $n$ can be mapped by an injective homomorphism to the general linear group $GL(d,\mathbf{F})$ over certain finite field $\mathbf{F}$, with dimension $d\leq n$. Each matrix element will be sorted in its own register, and then the register representing the group element is built from them. This encoding methods can be applied to generic discrete subgroups, including the largest discrete crystal-like $\bi$ subgroup of $SU(2)$ and $\vg$ of $SU(3)$. After reviewing the basics of group representation in \sect{encoding}, we introduce the block encoding methods. In \sect{primitive-gate}, we review the primary gates to implement group element operations and basic logic quantum gates set in constructing the primary gates. We then present the construction of primary gates based on the block encoding for the $\btt$, $\bi$ in \sect{unigates}. Resource requirements are estimated and compared to other digitization methods in \sect{resource}. Fidelities for the block encoding with quantum errors are analyzed in \sect{errors}, followed by the benchmarking of the inverse gates in \sect{expres}. In \sect{conclusion}, we present the conclusion and outlook. Appendices include an alternative constructions based on two's complement for block encoding.

\section{Discrete subgroup encodings} \label{sec:encoding}
For any encoding of discrete groups, the group elements $g$ are mapped to at least as many integers as $p=[0,|G|-1]$ where $|G|$ is the group dimension. Previous work focused on the \emph{ordered product} encoding which maps onto the integers the integer exponents $\{o_k\}$ of group generators $\lbrace\lambda_k\rbrace$ in a particular ordering:
\begin{align}
\label{eq:groupelement}
g_{\lbrace o_k \rbrace} =& \prod_k \lambda^{o_k}_k
\end{align}
where $0\leq o_k<O_k$. $O_k$ can be as large as the generator's order $\lambda_k^{O_k}=\mathbb{1}$ but often lower when redundancies occur e.g. $\lambda_1^{O_1}=\lambda_2$.  An integer mapping is then defined as
\begin{align}
\label{eq:intrep}
 p=o_k + O_{k}(o_{k-1}+O_{k-1}(\hdots+O_2(o_1+O_1o_0))).
\end{align}
With this, one may consider encoding $p$ onto quantum memory by decomposing $p$ via Eq.~(\ref{eq:intrep}) where $\{O_k\}$ are replaced by the dimensionality of the qudits $\{D_k\}$ available. Both $\btt$ and $\bo$ have been formulated in this way~\cite{Gustafson:2022xdt, Gustafson:2023kvd}. As an example, the group elements of $\btt$ can be represented as
\begin{equation}
    g = (-1)^m \mathbf{i}^n \mathbf{j}^o \mathbf{l}^{p}.
\end{equation}
where $\mathbf{i,j,k}$ are the unit quaternions and $\mathbf{l} = -(\mathbf{1 + i+ j+k})/2$. $m, n, o$ are 3 binary variables, while $p$ is tertiary, and thus can be encoded for example in a ($p=8$) quoctit and ($p=3$) qutrit or in 5 qubits with some forbidden states. A similar construction for $\bo$ could be realized with 6 qubits, or a qutrit with either 2 quoctits ($p=4$) or a qudecasexit ($p=16$) ~\cite{Gustafson:2023swx}. The corresponding qubit primitive gates to implement group inverse, product, trace and Fourier transformation has been worked out recently \cite{Gustafson:2022xdt, Gustafson:2023kvd}. While this encoding can be efficient in quantum memory, it is not a given that all finite groups can be encoding with ordered products.

Due to the potential limitations of the ordered product encoding, we consider a different encoding, block encoding, which instead maps the finite fields in the matrices representing $g$ to a set of qubits.
This encoding relies on the ability to represent the matrix elements themselves as valued only over a finite field. Here, we will consider how to encode $\btt$ and $\bi$ as examples.

The $\btt$ group is isomorphic to the special linear group $SL(2,3)$ — the group of all $2\times 2$ matrices with unit determinant over the three-element finite field $\mathbf{F}_3$ (which can be defined as the ring of integers modulo 3). The character table for $\btt$ is found at the bottom of Tab.~\ref{tab:chartab}. Using the isomorphism, $g$ can be represented by
\begin{eqnarray}
    g \in\left\{\begin{pmatrix}
a & b\\
c & d
\end{pmatrix}\bigg|\,a, b, c, d \in \mathbf{F}_3, ad-bc\equiv \,1\,\text{mod}(3) \right\}
\label{eq:btm}
\end{eqnarray}
This leads to another way of encoding $\btt$ group elements using four ternary variables $a, b, c, d=\{0,1,2\}$ as $\ket{g}=\ket{abcd}$ with a Hilbert space of $3^4 - 57=24$ states where 57 states are removed by the determinant constraints.

In this work, we encode each matrix element in \eq{btm} as a binary integer into two qubits, e.g. the non-negtive encoding with $\ket{0} = \ket{00}$, $\ket{1}=\ket{01}$, $\ket{2}  =\ket{10}$ and a forbidden state $\ket{11}$. 
Given the redundancy in this encoding, we can also use the two's complement encoding as $\ket{0}=\ket{00}, \ket{1}=\ket{01}$ and $\ket{2}  =\ket{11}$. In the $\btt$ case, the two's component encoding is the same as the Gray code~\cite{2004PhRvL..92q7902V,2020npjQI...6...49S,2021PhRvA.103d2405D,2021arXiv210308056C}. Alternatively, the matrix elements could each be encoding in a ($p=3$) qutrit without forbidden states. In this paper, we adopt the non-negative encoding, resulting in a 8-qubit group register while leaving optimizations of quantum resources among different encoding to the future as considerations to noise is crucial~\cite{Gustafson:2023swx}. 

\begin{table*}[htbp]
    \caption{Character Tables of (left) $\btt$ and (right) $\bi$ including a representative element in the given conjugacy class, where $\omega=e^{2\pi i/3}$, $\tau=(1+\sqrt{5})/2$, and $\sigma=(1-\sqrt{5})/2$.}
    \label{tab:chartab}
        \begin{tabular}{c|c|c|c|c|c|c|c}\hline
        $\btt$& $C_1$ & $C'_1$ & $C_4$ & $C'_4$ & $C''_4$ & $C'''_4$ & $C_6$ \\ \hline\hline
        Ord. & 1 & 2 & 3 & 3 & 6 & 6 & 4 \\ \hline
        $\chi_1$ & 1 & 1 & 1 & 1 & 1 & 1 & 1 \\
        $\chi_{1'}$ & 1 & 1 & $\omega$ & $\omega^2$ & $\omega$ & $\omega^2$ & 1 \\
        $\chi_{1''}$ & 1 & 1 & $\omega^2$ & $\omega$ & $\omega^2$ & $\omega$ & 1 \\
        $\chi_2$ & 2 & -2 & -1 & -1 & 1 & 1 & 0 \\
        $\chi_{2'}$ & 2 & -2 & $-\omega$ & $-\omega^2$ & $\omega$ & $\omega^2$ & 0 \\
        $\chi_{2''}$ & 2 & -2 & $-\omega^2$ & $-\omega$ & $\omega^2$ & $\omega$ & 0 \\
        $\chi_3$ & 3 & 3 & 0 & 0 & 0 & 0 & -1 \\[0.1cm]\hline
        $|g\rangle$ & \tiny $\begin{pmatrix}1&0\\0&1\end{pmatrix}$ & \tiny$\begin{pmatrix}2&0\\0&2\end{pmatrix}$ & \tiny$\begin{pmatrix}1&1\\0&1\end{pmatrix}$ & \tiny$\begin{pmatrix}1&2\\0&1\end{pmatrix}$ & \tiny$\begin{pmatrix}2&2\\0&2\end{pmatrix}$ & \tiny$\begin{pmatrix}2&1\\0&2\end{pmatrix}$ & \tiny$\begin{pmatrix}0&1\\2&0\end{pmatrix}$
    \end{tabular}\hspace{0.07cm}
    \begin{tabular}{c|c|c|c|c|c|c|c|c|c}\hline
        $\bi$ & $C_1$ & $C'_1$ & $C_{30}$ & $C_{20}$ & $C'_{20}$ & $C_{12}$ & $C'_{12}$ & $C''_{12}$ & $C'''_{12}$ \\\hline\hline
        Ord.& 1 & 2 & 4 & 3 & 6 & 5 & 10 & 5 & 10 \\\hline
        $\chi_1$ & 1 & 1 & 1 & 1 & 1 & 1 & 1 & 1 & 1 \\
        $\chi_2$ & 2 & -2 & 0 & -1 & 1 & $-\tau$ & $\tau$ & $-\sigma$ & $\sigma$ \\
        $\chi_{2'}$ & 2 & -2 & 0 & -1 & 1 & $-\sigma$ & $\sigma$ & $-\tau$ & $\tau$ \\
        $\chi_3$ & 3 & 3 & -1 & 0 & 0 & $\tau$ & $\tau$ & $\sigma$ & $\sigma$ \\
        $\chi_{3'}$ & 3 & 3 & -1 & 0 & 0 & $\sigma$ & $\sigma$ & $\tau$ & $\tau$ \\
        $\chi_4$ & 4 & 4 & 0 & 1 & 1 & -1 & -1 & -1 & -1 \\
        $\chi_5$ & 5 & 5 & 1 & -1 & -1 & 0 & 0 & 0 & 0 \\
        $\chi_{4'}$ & 4 & -4 & 0 & 1 & -1 & -1 & 1 & -1 & 1 \\
        $\chi_6$ & 6 & -6 & 0 & 0 & 0 & 1 & -1 & 1 & -1 \\\hline
        $|g\rangle$ & \tiny$\begin{pmatrix}1&0\\0&1\end{pmatrix}$ & \tiny$\begin{pmatrix}4&0\\0&4\end{pmatrix}$ & \tiny$\begin{pmatrix}0&1\\4&0\end{pmatrix}$ & \tiny$\begin{pmatrix}0&1\\4&4\end{pmatrix}$ & \tiny$\begin{pmatrix}0&4\\1&1\end{pmatrix}$ & \tiny$\begin{pmatrix}1&1\\0&1\end{pmatrix}$ & \tiny$\begin{pmatrix}4&4\\0&4\end{pmatrix}$ & \tiny$\begin{pmatrix}1&2\\0&1\end{pmatrix}$ & \tiny$\begin{pmatrix}4&3\\0&4\end{pmatrix}$ \\
    \end{tabular}
\end{table*}

The block encoding can be extended to other groups. The $\bi$ group is isomorphic to the special linear group $SL(2,5)$ — the group of all $2\times 2$ matrices over the finite field $\mathbf{F}_5$ with unit determinant. The character table for $\bi$ is found at the bottom of Tab.~\ref{tab:chartab} and the group element can be represented by
\begin{eqnarray}  
g \in\left\{\begin{pmatrix}
a & b\\
c & d
\end{pmatrix}\bigg|\,a, b, c, d \in \mathbf{F}_5, ad-bc\equiv 1 \text{ mod}(5)\right\}
\label{eq:bi}
\end{eqnarray}
We can encode the $\bi$ group elements with four quinary variables $a, b, c, d=\{0,1,2,3,4\}$, and thus the matrix elements can be represented with three qubits with forbidden states $\ket{101},\ket{110},\ket{111}$, and the group element with 12 qubits. Alternatively, the matrix elements could be encoding into ($p=5$) ququints without forbidden states.  In this work, we derive the primitive gates of $\bi$ with a 12-qubit group register.

It is worth noting that the largest crystal-like subgroup of $SU(3)$ -- $\vg$ -- is isomorphic to a subgroup of $GL(3,4)$ 
and can be encoded with nine 9 ququarts or 18 qubits, though further investigations are needed to implement the quantum gates that can project the Hilbert space to its $\vg$ subspace.

\section{Primitive gate overviews}\label{sec:primitive-gate}
Quantum circuits for pure gauge theories can be constructed out of a set of primitive gates~\cite{Lamm:2019bik} acting on one or more group element registers.  The necessary gates for simulation and extraction of observables are:
\begin{itemize}
    \item the inverse gate: $\mathfrak{U}_{-1}|g\rangle=\left|g^{-1}\right\rangle$, which computes in-place the inverse of a group element,
    \item the trace gate: $\mathfrak{U}_{\operatorname{Tr}}(\theta)|g\rangle=e^{i \theta \operatorname{Re} \operatorname{Tr} g}|g\rangle$, which introduces a phase based on the trace of a group element in a particular representation and weighted by a coupling $\theta$ which can depend on the Hamiltonian and approximation used in time evolution, 
    \item the left multiplication gate: $\mathfrak{U}_{\times}|g_i\rangle|g_j\rangle|{\rm anc}\rangle=|g_i\rangle|g_j\rangle|g_ig_j\rangle$. In this work, we consider a novel definition of $\mathfrak{U}_{\times}$ which stores the result in an ancilla group register. The right multiplication gate, if desired, can be defined via the $\mathfrak{U}_{-1}$, and $\mathfrak{U}_{\times}$,
    \item the group Fourier gate: $
\mathfrak{U}_{F}\sum_{g}f(g)\left|g\right> = \sum_{\rho}\hat{f}(\rho)_{ij}\left|\rho, i, j\right>$ with $\hat f$ denoting the Fourier transform of $f$.
\end{itemize}

For qubit-based computers, we need 8 qubits to encode a $\btt$ register and 12 qubits for $\bi$. We will present the primitive gates construction using five entangling gates: the two qubit CNOT and SWAP, the three qubit CSWAP (known as the Fredkin gate), and the multi-controlled phase ${\rm C}^n{\rm P(\phi)}$ and C$^n$NOT quantum gates. The special case of C$^2$NOT is commonly called the Toffoli gate.  The first three implement the operations:
\begin{align}
    \text{CNOT}\ket{q_1}\ket{q_2} &= \ket{q_1}\left|q_1\oplus q_2\right>\\
    \text{SWAP}\ket{q_1} \ket{q_2} &= \ket{q_2} \ket{q_1}\\
     {\rm CSWAP}\ket{q_1}\ket{q_2}\ket{q_3}&=\ket{q_1}\ket{q_3}\ket{q_2}.
\end{align}
For the multi-controlled gates, which apply and operation to one qubit based on the state of $n-1$ others we have
\begin{align}
   {\rm C}^n{\rm P(\phi)}\ket{q_1}\dots\left|q_n\right>&=\ket{q_1}\dots e^{i\phi q_1\dots q_{n-1}}\left|q_n\right>\\
    {\rm C}^n{\rm NOT}\ket{q_1}\dots\left|q_n\right>&=\ket{q_1}\dots\left|q_n\oplus q_1\dots q_{n-1}\right>.
\end{align}

\section{Primitive Gates for \texorpdfstring{$\btt$}{BT} and \texorpdfstring{$\bi$}{BI}}\label{sec:unigates}
To construct the primitive gates based on block encodings, we adopt the following conventions.  We use parentheses to denote operations that return values with mod $n$ applied for $SL(2,n)$ group. The values of matrix element are the remainders mod $n$. A feature of the block encoding is that the primitive gates for $\btt$ and $\bi$ are quite similar, with the difference mainly arising from the modular arithmetic.  Therefore we will consider the two groups in parallel as we construct their gates.

\subsection{Inverse Gate}
\begin{figure*}[htbp]
  \centering
   \begin{minipage}[h]{0.12\linewidth}
	\vspace{0pt}
	\includegraphics[width=\linewidth]{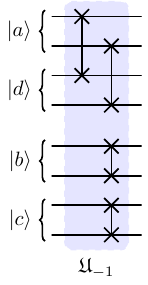}
   \end{minipage}
   ~\begin{minipage}[h]{0.7\linewidth}
	\vspace{0pt}
	\includegraphics[width=\linewidth]{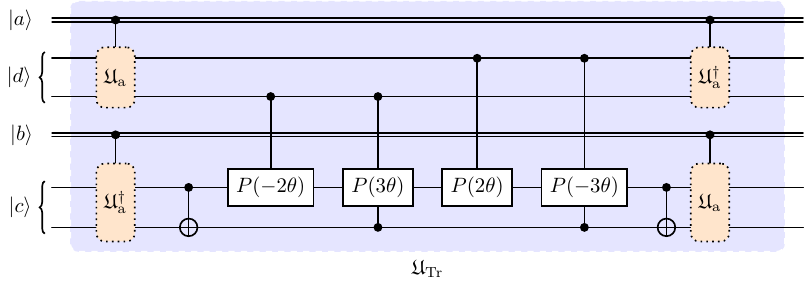}
   \end{minipage}
  \caption{Inverse gate $\mathfrak{U}_{-1}$ (left) and Trace gate $\mathfrak{U}_{\rm Tr}(\theta)$ (right) for $\btt$ group.}
    \label{fig:btprime1}
\end{figure*}

\begin{figure*}[htbp]
  \centering
   \begin{minipage}[h]{0.12\linewidth}
	\vspace{0pt}
	\includegraphics[width=\linewidth]{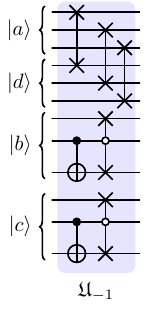}
   \end{minipage}
~\begin{minipage}[h]{0.85\linewidth}
	\vspace{0pt}
 \includegraphics[width=\linewidth]{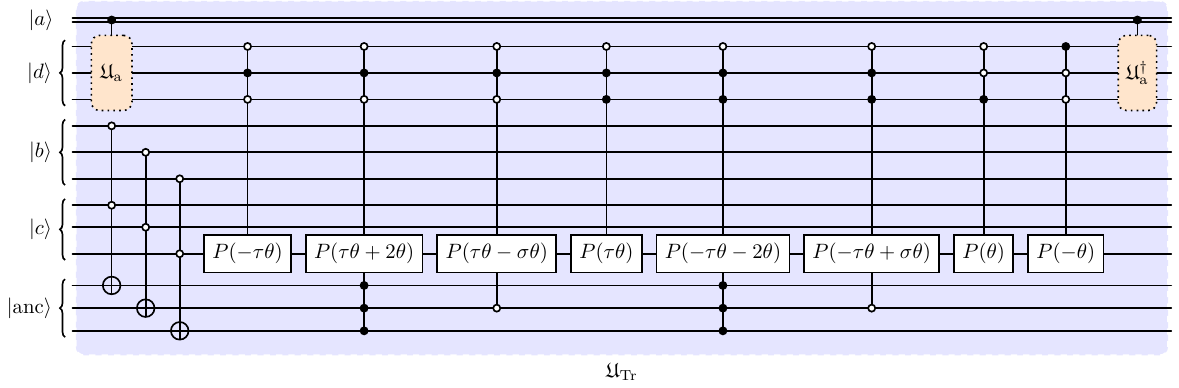}
   \end{minipage}
  \caption{Inverse gate $\mathfrak{U}_{-1}$ (left) and trace gate $\mathfrak{U}_{\rm Tr}$ (right) for $\bi$ group.}
    \label{fig:primeBI}
\end{figure*}

For the construction of $\mathfrak{U}_{-1}$, we note that the inverse $g^{-1}$ is given in terms of the matrix elements of $g$ by 
\begin{eqnarray}
    g^{-1} = \begin{pmatrix}
d & -b\\
-c & a
\end{pmatrix},
\label{eq:btmin}
\end{eqnarray}
From this, we see that the inverse operation corresponds to swapping the values of $a,d$ and flipping the sign of $b,c$. Using this, the $\mathfrak{U}_{-1}$ circuits can be derived and are presented in the left of \fig{btprime1} and \fig{primeBI} for $\btt$ and $\bi$ respectively. 

\subsection{Trace Gate}

Our interest here is in the traces of each group elements in the faithful representation, which are denoted by $\chi_2$ in \tab{chartab}. With one representative element in the given conjugacy class in \tab{chartab}, one can determine the other elements in the class. This enables a derivation of the rules for obtaining $\tr(g)$ in the blocking encoding.  For $\btt$, these rules are
\begin{eqnarray}\label{eq:tr}
    \tr(g) =\begin{dcases}
         2 & {\rm if~} (a+d)=2 \,\&\, (c-b)=0 \\
         1 & {\rm if~} (a+d)=1 \,\&\, (c-b)\neq 0 \\
         0 & {\rm if~} (a+d)=0 \\
        -1 & {\rm if~} (a+d)=2 \,\&\, (c-b)\neq 0 \\
        -2 & {\rm if~} (a+d)=1 \,\&\, (c-b)=0 
    \end{dcases}
\end{eqnarray}
While for $\bi$, they are found to be
\begin{eqnarray}
    \tr(g) =\begin{dcases}
    0 & {\rm if~} (a+d)=0 \\
    1 & {\rm if~} (a+d)=1 \\
    -1 & {\rm if~} (a+d)=4 \\
    2 & {\rm if~} (a+d)=2 \,\&\, (c^2+b^2)=0 \\
    -\tau & {\rm if~} (a+d)=2 \,\&\, (c^2+b^2)=1{\rm ~or~}2 \\
    -\sigma & {\rm if~} (a+d)=2 \,\&\, (c^2+b^2)=3{\rm ~or~}4 \\
    -2 & {\rm if~} (a+d)=3 \,\&\, (c^2+b^2)=0 \\
    \tau & {\rm if~} (a+d)=3 \,\&\, (c^2+b^2)=1{\rm ~or~}2 \\
    \sigma & {\rm if~} (a+d)=3 \,\&\, (c^2+b^2)=3{\rm ~or~}4 \\
    \end{dcases}.
\end{eqnarray}

\begin{figure*}[htbp]
  \centering
   \begin{minipage}[h]{0.15\linewidth}
	\vspace{0pt}
	\includegraphics[width=\linewidth]{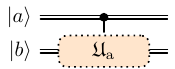}
   \end{minipage}
  = \begin{minipage}[h]{0.3\linewidth}
	\vspace{0pt}
	\includegraphics[width=\linewidth]{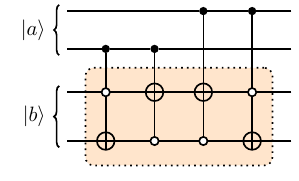}
   \end{minipage}
~~~~\begin{minipage}[h]{0.15\linewidth}
	\vspace{0pt}
 \includegraphics[width=\linewidth]{add1}
   \end{minipage}
  = \begin{minipage}[h]{0.3\linewidth}
	\vspace{0pt}
	\includegraphics[width=\linewidth]{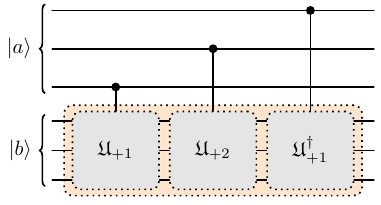}
   \end{minipage}
  \caption{Implementation of $\mathfrak{U}_{\rm a}$ for $\btt$ (left) and $\bi$ group(right).}
    \label{fig:addsub}
\end{figure*}

\begin{figure}[htbp]
  \centering
   \begin{minipage}[h]{0.75\linewidth}
	\vspace{0pt}
	\includegraphics[width=\linewidth]{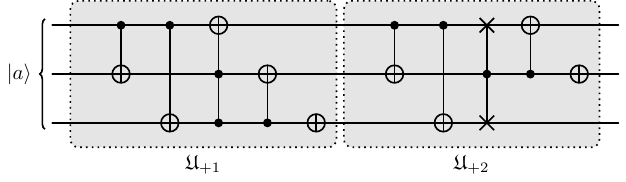}
   \end{minipage}
  \caption{Elementary operations $\mathfrak{U}_{\rm +1}$ and $\mathfrak{U}_{\rm +2}$ that add one and two to a single matrix element, respectively for the $\bi$ group.}
    \label{fig:addbi12}
\end{figure}

By inspecting these rules, one notices that to realize $\mathfrak{U}_{\rm Tr}(\theta)$, we first implement an addition operation $\mathfrak{U}_{\rm a}\ket{a}\ket{b}=\ket{a}\ket{b\oplus a}$. These are constructed for $\btt$ and $\bi$ in \fig{addsub}. In the case of $\bi$, it is useful to further define subroutines $\mathfrak{U}_{\rm +n}$ for $n=1,2$ which increment the matrix element $\ket{a}$ to $\ket{a+n}$ as shown in \fig{addbi12}. The operation of $(b - a)$ can be implemented using $\mathfrak{U}^\dagger_{\rm a}$. With $\mathfrak{U}_{a}$, we can construct $\mathfrak{U}_{\rm Tr}(\theta)$, as shown in the right of \fig{btprime1} and \fig{primeBI} where $\theta$ depends on couplings and approximations in the time evolution.

\begin{figure*}[htbp]
  \centering
   \begin{minipage}[h]{0.15\linewidth}
	\vspace{0pt}
	\includegraphics[width=\linewidth]{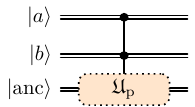}
   \end{minipage}
  = \begin{minipage}[h]{0.3\linewidth}
	\vspace{0pt}
	\includegraphics[width=\linewidth]{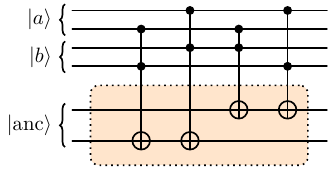}
   \end{minipage}
~~~\begin{minipage}[h]{0.13\linewidth}
	\vspace{0pt}
	\includegraphics[width=\linewidth]{prod1}
   \end{minipage}
  = \begin{minipage}[h]{0.34\linewidth}
	\vspace{0pt}
	\includegraphics[width=\linewidth]{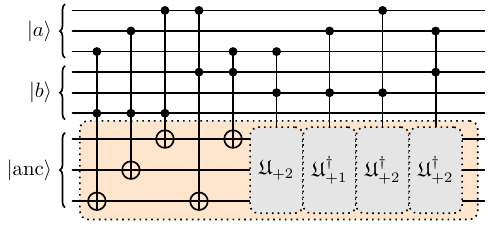}
   \end{minipage}
  \caption{Implementation of $\mathfrak{U}_{\rm p}$ for $\btt$ (left) and $\bi$ group(right).}
    \label{fig:prodsub}
\end{figure*}

\subsection{Multiplication Gate}
Moving on to $\mathfrak{U}_{\times}$, the multiplication of two group elements $g_i$ and $g_j$ in terms of the matrix elements is
\begin{equation}
\begin{pmatrix}
a_i & b_i\\
c_i & d_i
\end{pmatrix} \begin{pmatrix}
a_j & b_j\\
c_j & d_j
\end{pmatrix} = 
\begin{pmatrix}
a_ia_j + b_i c_j & a_i b_j+ b_i d_j\\
c_i a_j+ d_i c_j & c_i b_j+ d_i d_j
\end{pmatrix}
\label{eq:multi}
\end{equation}

\begin{figure}
\centering    
\includegraphics[width=\linewidth]{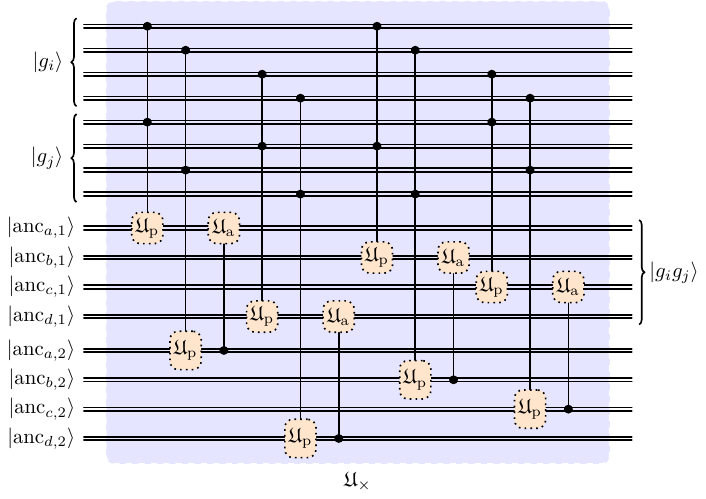}
\caption{Multiplication gate $\mathfrak{U}_{\times}$ for $\btt$ group and $\bi$ group.}
    \label{fig:btmulti}
\end{figure}

The multiplication of group elements can thus be built from blocks of $\mathfrak{U}_{\rm a}$ and a new subroutine $\mathfrak{U}_{\rm p}$ which calculate the product of two elements (See \fig{prodsub}). In order to leave the matrix elements intact, ancillary qubits store the product. One ancillary register will be required to store each matrix element in \eq{multi}. Nevertheless, to optimise the parallelization, we actually introduce one ancillary register for each product in \eq{multi}, which doubles the ancillary qubits required. The group multiplication circuit $\mathfrak{U}_\times$ is subsequently constructed as in \fig{btmulti}.

\subsection{Fourier Gate}
To reduce the quantum gates required for implmenting Fourier gate, we can project both $\btt$ and $\bi$ group to the subspace satisfying the determinate constraints using circuit $\mathfrak{U}_{proj}$ and ancillary qubits, which will reduce the encoding from 8 and 12 qubits to 6 and 9 qubits, respectively. $\mathfrak{U}_{proj}$ is implemented with 4 CNOT gates, 3 CSWAP gates and 8 Tofolli gates for $\btt$ group. For the $\bi$ group, $\mathfrak{U}_{proj}$ requires 8 CNOT gates, 3 CSWAP gates, 8 Toffoli gates and 2 $\mathfrak{U}_{\rm p}$ gates. With such projection, transforming to the Fourier basis can be decomposed using the \textsc{Qiskit} transpiler. The resource required for Fourier gate implemented this way for both groups are shown in \tab{uf}.

\begin{table}[]
    \centering
    \begin{tabular}{c|c| c| c| c }
    \hline
    &\multicolumn{4}{|c}{$\mathfrak{U}_F$} \\  \cline{2-5}
               & $R_X$ & $R_Y$ & $R_Z$ & CNOT\\ 
               \hline
       $\btt$  & 186& 2052 &3491&1941\\\hline
        $\bi$ & 10743& 131879 & 223044& 125919 \\\hline
    \end{tabular}
    \caption{$\mathfrak{U}_F$ decompositions for $\btt$ and $\bi$ group.}
    \label{tab:uf}
\end{table}
$\mathfrak{U}_F$ is usually the most costly primitive gate for quantum simulations. Future direction should consider the fast Fourier transformation in \cite{moore2003generic} where sub-exponential growth $\mathcal{\exp(\sqrt{\log|G|})}$ of the circuit depth in the group size is possible. In particular, it should be noticed that the Fourier transformation of $\bi$ group can be built upon that of $\btt$ group using the following natural tower:
\begin{equation}
    \bi > \btt > \mathbb Z_6 > \mathbb Z_3 > \mathbb Z_1=\{1\}.
\end{equation}
which could simplify the constructions for the $\bi$ group once the construction for $\btt$ is realized. It will be valuable to realize such fast Fourier transformation and compare the resources required with our brutal direct decomposition, which we will leave for the future work.

With these four primitive gates, it is possible to perform a resource estimate and compare to other implementations of $SU(2)$. This will be done in \sect{resource}.  Additionally, the primitive gates for the two's component encoding are discussed in \seca{bt2}.  There we find that total gate costs are similar, but given the different distribution of qubit states, they are anticipated to have different noise robustness.

\section{Resource Requirements}
\label{sec:resource}
As T gate counts are known to require costly encodings for error correction, it is usually used in fault-tolerent resource analysis \cite{Nielsen-qec, Kitaev_1997}. We estimate the T gates required for simulating both $\btt$ or $\bi$ group with the block encoding method. The Toffoli gate requires 7 T gates \cite{Nielsen-qec}; ${\rm C}^n{\rm NOT}$ can be constructed with $(2[\log_2(n+1)] - 1)$ toffoli gates and $n-2$ dirty ancilla qubits for $n > 2$ \cite{Nielsen-qec}. ${\rm C}^n{\rm SWAP}$ require the same number of T gates as ${\rm C}^{n+1}{\rm NOT}$, as it can be decomposed to ${\rm C}^{n+1}{\rm NOT}$ and CNOTs using the symmetric decomposition \cite{M_Q_Cruz_2024}. 
Approximating $R_Z$ gates to the precision of $\epsilon$ will require $1.15 \log_2(1/\epsilon)$ T gates \cite{Gustafson:2023kvd}, while $R_X$ and $R_Y$ can be implemented with at most three $R_Z$ gates. ${\rm C}{\rm P(\phi)}$ requires 8 T gate and one $R_Z$ gate with one clean ancillary qubit \cite{Kim2018EfficientDM}, and ${\rm C}^n{\rm P(\theta)}$ can be decomposed to $2(n-1)$ Toffoli gates and one ${\rm C}{\rm P(\theta)}$ with $n-1$ clean ancilla \cite{QC-infor}.
The estimated number of T gates to realize each primary gates are listed in \tab{t-count}. 
\begin{table}[]
    \centering
    \begin{tabular}{c|c|c}
    \hline
    gate & T ($\btt$) & T ($\bi$) \\
    \hline
        $\mathfrak{U}_{-1}$ & 0 & 14\\
    \hline
        $\mathfrak{U}_{\rm Tr}$& 172 + 4.6~$\log_2(1/\epsilon)$ & 666 + 9.2~$\log_2(1/\epsilon)$ \\ 
    \hline
        $\mathfrak{U}_{\times}$& 336 & 3640\\
    \hline
        $\mathfrak{U}_{F}$&$11735.8\log_2(1/\epsilon)$ &$748547\log_2(1/\epsilon)$\\
    \hline
    \end{tabular}
    \caption{Number of T gates required to realize the primary gates.}
    \label{tab:t-count}
\end{table}

We compare the resources required for the calculation of the viscosity as that in \cite{kan2022lattice}. The total T gate count for certain Hamiltonian $H$ is given by $N^{H}_T=C^H_T\times d L^d N_t$ for a $d$ spatial lattice simulated for a time $t=N_t\delta t$ \cite{Gustafson:2023kvd}, where $C^H_T$ is the average number of T gates require per link per $\delta t$, and $d L^d$ is the total number of links. We first consider the simulation of the $\btt$ group. With the primitive gates per link per $\delta t$ listed in \cite{Gustafson:2023kvd}, we get
\begin{align}
    C^{H_{\rm KS}}_{T}&=2102(d-1) + (23469.2 + 2.3  d)\log_2\frac{1}{\epsilon}\\
    C^{H_I}_{T}&=8994 d-7650 + (46936.1 + 6.9  d)\log_2\frac{1}{\epsilon},
\end{align}
using the Kogut-Susskind Hamiltonian $H_{\rm KS}$ and the improved Hamiltonian $H_I$ studied in \cite{Carena:2022kpg}, respectively. With this, the total synthesis error $\epsilon_T$ can be estimated as the sum of $\epsilon$ from each $R_Z$:
\begin{align}
    \epsilon^{H_{KS}}_T&=2 (10204 + d)d L^d N_t\times \epsilon\\
    \epsilon^{H_{I}}_T &=2(20407 + 3d)d L^d N_t \times\epsilon.
\end{align}
For simulating the $\bi$, the cost increase moderately:
\begin{align}
    C^{H_{\rm KS}}_{T}&=22215(d-1) + (1497090 + 4.6  d)\log_2\frac{1}{\epsilon}\\
    C^{H_I}_{T}&=95793 d-81205 + (2994170 + 13.8  d)\log_2\frac{1}{\epsilon},
\end{align}
and total synthesis errors:
\begin{align}
    \epsilon^{H_{KS}}_T&=4  (325454 + d)d L^d N_t\times \epsilon\\
\epsilon^{H_{I}}_T &=12  (216969 + d)d L^d N_t \times\epsilon.
\end{align}

To calculate the shear viscosity with the total synthesis error $\epsilon_T = 10^{-8}$ on a $d=3$ lattice with $L^3 = 10^3$ for $N_t = 50$, finding
$N^{H_{\rm KS}}_T=2.1\times 10^{11}$ and
$N^{H_I}_T=4.2\times 10^{11}$. For the ordered product method in simulating $\btt$ group \cite{Gustafson:2022xdt}, the total number of T gates are estimated as 1.1$\times10^{11}$ for $H_{\rm KS}$ and 4.1$\times10^{11}$ for $H_I$. We note that $H_{\rm KS}$ is inadequate to reach the scaling regime with $\btt$, but for the case of $H_I$, the T gate cost is only $\sim 3\%$ higher -- and in both cases dominated by $\mathfrak{U}_F$. With only 60\% higher qubit costs, we conclude that this encoding is comparable to ordered product on these simple metrics, and further analysis of other metrics like noise robustness should be undertaken. For the larger $\bi$ group -- where no ordered product encoding currently exists -- the costs increase to $N^{H_{\rm KS}}_T=1.4\times 10^{13}$ and $N^{H_I}_T=2.9\times 10^{13}$ T gates using $H_{KS}$ and $H_I$, respectively. But given $\beta_f\gg\beta_s$ for $\bi$, the increase in gate costs may be an acceptable trade-off due to the reduced systematic error from digitization.

We can also compare the the block encoding of discrete subgroups to other digitization methods, such as LSH formalism, in which the T gate counts has been estimated for $d=1$ in \cite{Davoudi_2023}. For comparison, we take the finite arithmetic precision errors to be smaller than the synthesis error $\epsilon_T=10^{-8}$ by taking minimally $n=7$ steps and $m = 42$ bits to evaluate the inverse-square-root functions using Newton's method. We choose 5 and 7 qubits to encode one group register in LSH to match the minimal number of qubits that can encode $\btt$ and $\bi$ group, respectively. With $L = 10$, LSH will require in trotter step using singular-value-decomposition roughly $3.9\times 10^7$ and $4.2\times 10^7$ T gates for the 5 qubits and 7 qubits case, respectively, while for the blocking encoding, $\btt$ and $\bi$ requires $1.3\times10^7$ and $8.0\times 10^8$ T gates. Given this similarity in resource costs, quantifying the systematic and statistical errors in digitization methods becomes an important concern. In the next section, we consider the relative resilience of the block encoding versus the ordered producted encoding.  Additionally, a comprehensive comparision to other forms of approximating the time evolution in LGT~\cite{Shaw:2020udc,Rhodes:2024zbr} are desirable, but understanding the theoretical LGT errors must be understood for a complete comparison.

\section{Resilience to Quantum Errors}\label{sec:errors}
In this section, we investigate the resilience to errors for a single register using the block encoding and the ordered product encoding, considering $\btt$. Given that redundant degrees of freedom are introduced in both encodings, one can utilize these redundancies in QEC~\cite{PhysRevLett.125.240405,Bonati:2021vvs,Bonati:2021hzo,Gustafson:2023swx,Stannigel:2013zka,Tran:2020azk,Lamm:2020jwv,Stryker:2018efp,Rajput2023npj,Mathew:2022nep,VanDamme:2021njp,Halimeh:2022mct, Carena:2024dzu}. In the following, we derive the error rate threshold~\cite{Carena:2024dzu} below which the block encoding would provide higher fidelity than the ordered product one for a single register as more redundancies are introduced in the block encoding. We will take the bit-flip error as an example and consider the fidelity as the one averaging over all group elements. Error mitigation using post-selection remove trials affected by detectable errors. For the error channel with only one bit-flip error $\mathcal{N}_i$ affecting qubit $i$ at an error rate of $\epsilon$, we count the number of $\mathcal{N}_i$ that transform the group element into forbidden state, which are detectable errors. Using the ordered product encoding, there are in total 16 detectable $\mathcal{N}_i$ for all 24 group elements, while for the block encoding, 160 $\mathcal{N}_i$. The lower bound for the fidelity is the probability of no errors after removing trials affected by these detectable one bit-flip errors:
\begin{eqnarray}
    \mathcal{F}^{{\rm ps}}_{\rm BE} &\geq& \frac{(1-\epsilon)^8}{1-160/24(1-\epsilon)^7 \epsilon},\notag\\
    \mathcal{F}^{{\rm ps}}_{\rm OPE} &\geq& \frac{(1-\epsilon)^5}{1-16/24(1-\epsilon)^4\epsilon}.
\end{eqnarray}
For logical error rates $\epsilon \lesssim 0.1$, we found the lower bound of the fidelity for the  current encoding methods can be higher. As post-selections require resources that are exponential \cite{cai2023quantum} in system size, we also consider correcting one bit-flip error to reduce the effects of quantum errors.
The correctable one bit-flip errors transform the group element to a forbidden state that cannot be transformed from other group element via any one bit-flip error (Knill-Laflamme condition \cite{KnillQECC97}). Under this condition, when a forbidden state is seen, the error channel can be inferred and corrected. Using the ordered product encoding, there are in total 0 correctable $\mathcal{N}_i$ for all 24 group elements, while for the block encoding, 80 $\mathcal{N}_i$. Correcting trials affected by correctable $\mathcal{N}_i$, we obtain the lower bound of the averaged fidelity as the probability with no errors or only correctable one bit-flip errors:
\begin{eqnarray}
    \mathcal{F}^{\rm cor}_{\rm BE} &\geq& (1-\epsilon)^8 +\frac{80}{24}(1-\epsilon)^7\epsilon,\notag\\
    \mathcal{F}^{\rm cor}_{\rm OPE} &\geq& (1-\epsilon)^5.
\end{eqnarray}
We observe that in this situation for $\epsilon \lesssim 0.1$, the lower bound of the fidelity for the current encoding methods can be mildly higher.

\section{Experimental Results}
\label{sec:expres}
\begin{figure*}[htbp]
    \centering
\begin{minipage}[h]{0.67\linewidth}
      \vspace{0pt}
\includegraphics[width=\linewidth]{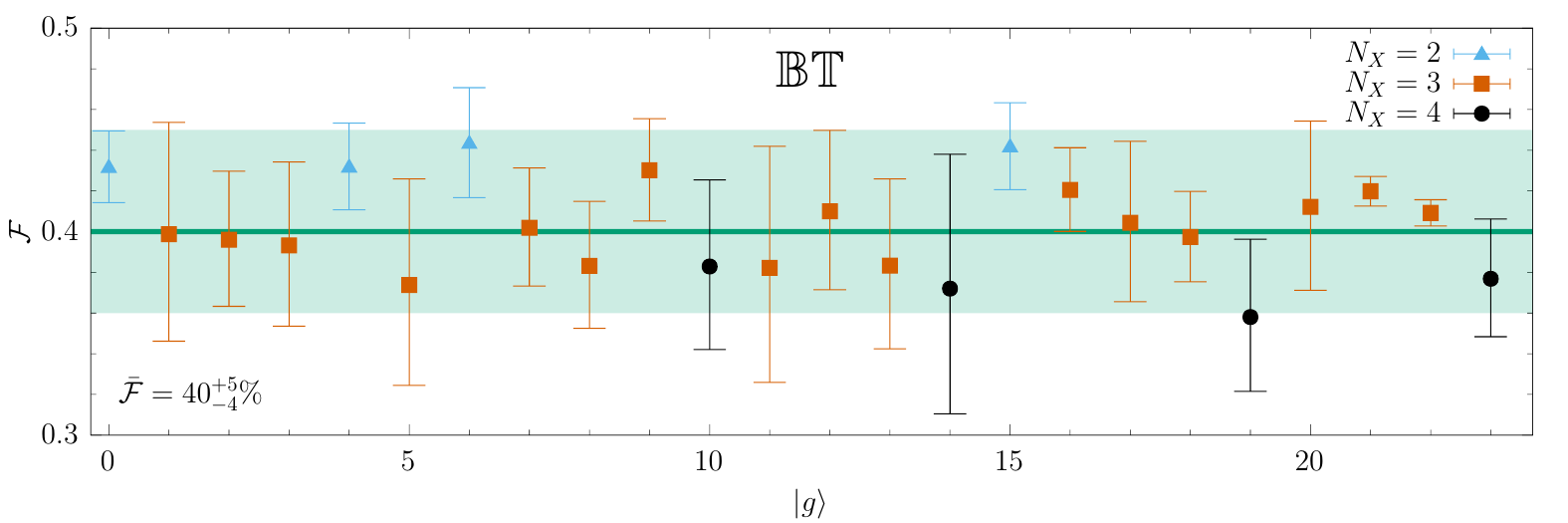} 
     \end{minipage}
     ~\begin{minipage}[h]{0.15\linewidth}
      \vspace{-10pt}      \includegraphics[width=\linewidth]{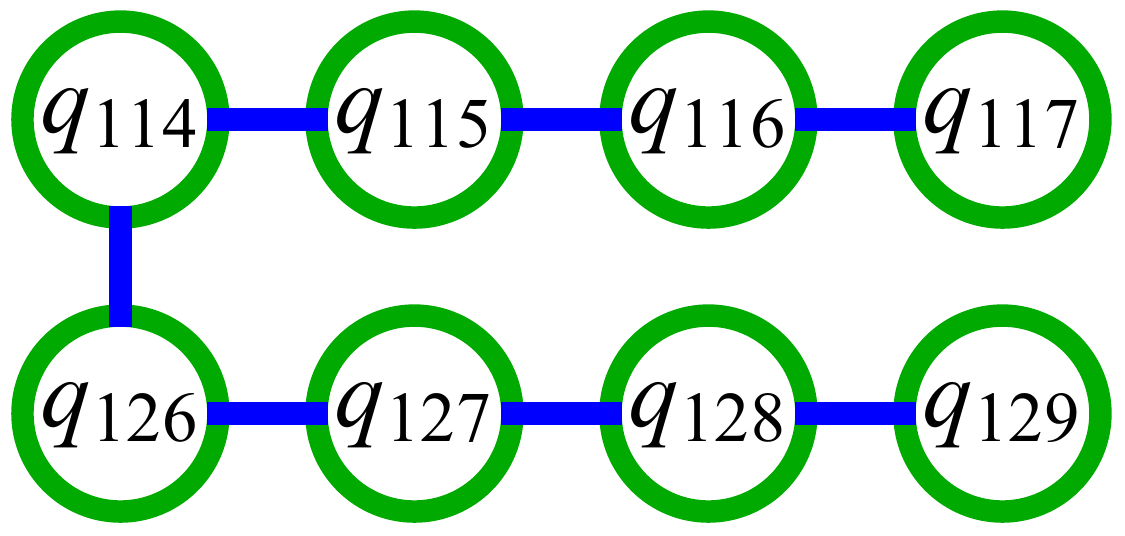} 
     \end{minipage}
     ~\begin{minipage}[h]{0.15\linewidth}
      \vspace{-8pt}
\includegraphics[width=\linewidth]{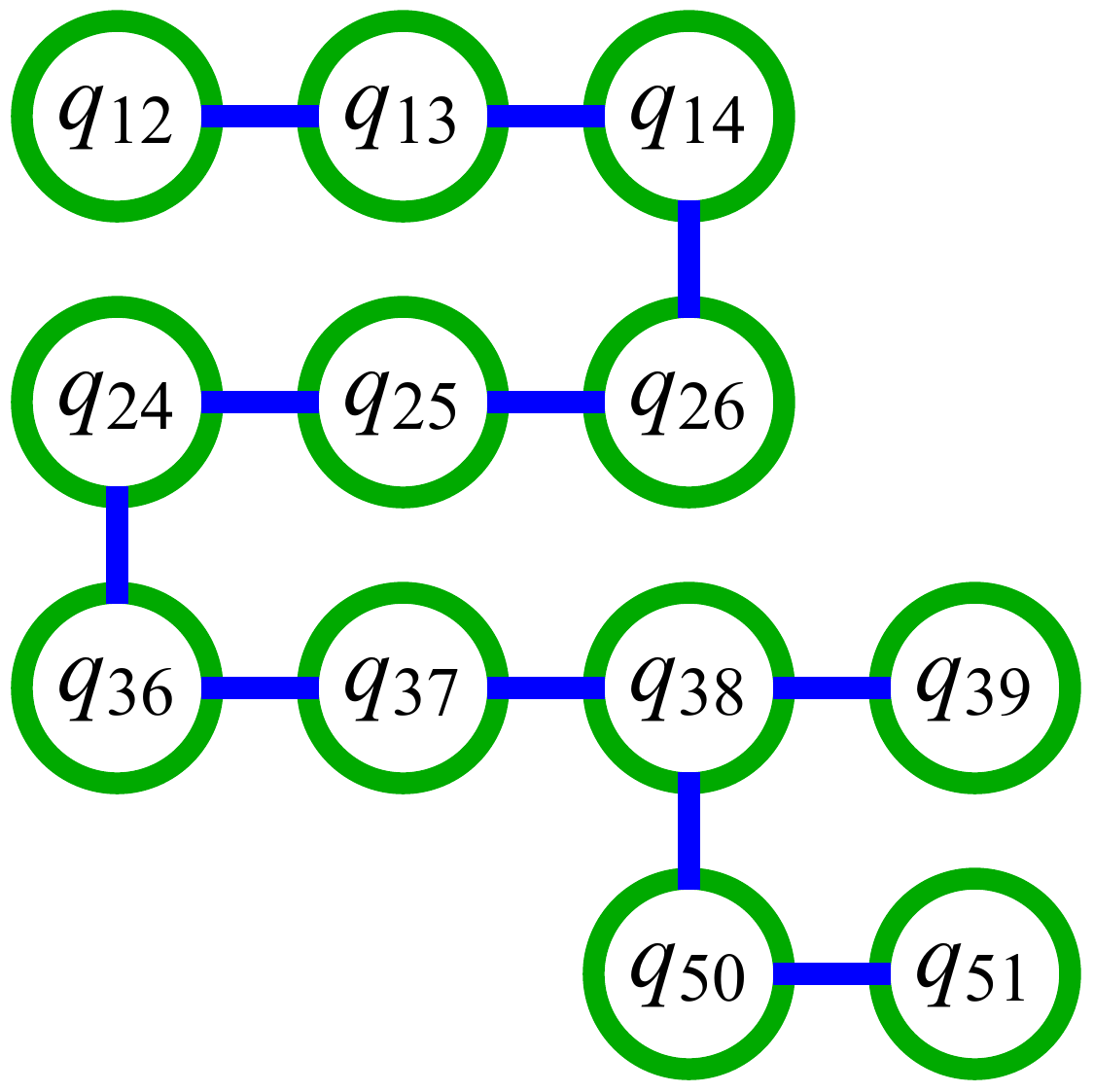} 
     \end{minipage}\\
      \vspace{0pt}
\includegraphics[width=\linewidth]{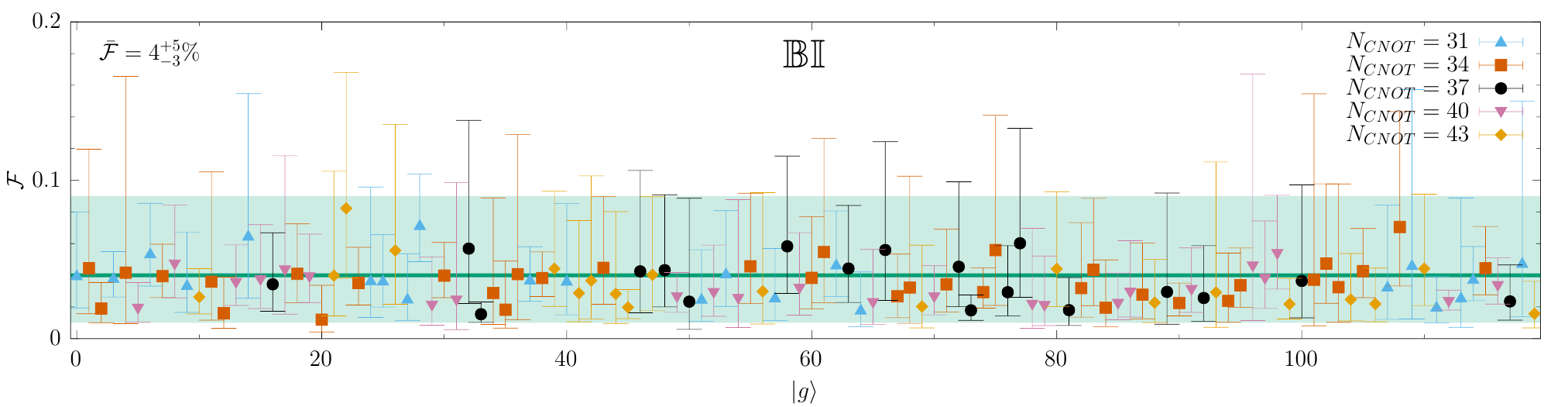} 
    \caption{Fidelity of $\mathfrak{U}^{\rm trans}_{-1}$ gate of \texttt{Baiwang} for each group element $|g\rangle = |abcd\rangle$ of $\btt$ and $\bi$, which is labeled by the lexicographic order in the range $\{0,|G|-1\}$. The averaged fidelity $\bar{\mathcal{F}}$ over all group elements are also shown. On the top right, the qubit graph on \texttt{Baiwang} used to represent $\ket{g}$ for (left) $\btt$ and (right) $\bi$ respectively.}
    \label{fig:fidelity}
\end{figure*}
Using the block encoding method, we benchmark the fidelity of the inverse gate for $\btt$ and $\bi$ group on the Quafu platform \cite{bib40}. Given the availability, we used the \texttt{Baiwang} quantum real machine on Quafu which has 144 qubits and 136 available.
The qubits are arranged in a $12\times 12$ lattice, where qubits in each row are connected adjacently and adjacent rows are connected by only 2-4 qubit connections \cite{bib40}. 8 qubits and 12 qubits are selected for the simulation of $\btt$ and $\bi$ group, respectively, with their positions and connectivity shown in~\fig{fidelity}. 
Starting from state $|\psi_0\rangle = |0\rangle^{\otimes n_q}$ with $n_q = 8(12)$ for $\btt(\bi)$ group, we can prepare the initial state $|g\rangle$ with circuit $\mathfrak{U}_g$: $|g\rangle = \mathfrak{U}_g |\psi_0\rangle$. For single states, $\mathfrak{U}_g$ is simply a tensor product of X gates used to initialize some qubits to $|1\rangle$. Adapting to the connectivity of the qubits chosen on \texttt{Baiwang} quantum chip, $\mathfrak{U}_{-1}$ requiring 12 (29) CNOT gates\footnote{We have decomposed the Fredkin gate to 7 CNOTs~\cite{M_Q_Cruz_2024}}) is transpiled to $\mathfrak{U}^{\rm trans}_{-1}$ with 18 (31-43) CNOTs for $\btt (\bi)$ where the number of CNOTs is counted from the transpiled circuits. We subsequently apply 
$\mathfrak{U}^{\rm trans}_{-1}$ to $|g\rangle$, resulting in the final state $|g'\rangle$. The fidelity is defined as:
\begin{equation}
    \mathcal{F} = |\langle g^{-1}|g'\rangle|^2 =  |\langle g^{-1}|\mathfrak{U}^{\rm trans}_{-1}\mathfrak{U}_g |\psi_0\rangle|^2,
\end{equation}
which is the probability of measuring the correct $|g^{-1}\rangle$. 

We show the fidelity for each $g$ in \fig{fidelity} for both $\btt$ and $\bi$ group calculated from 5 runs, with $N=5000$ shots for each run. The group element are labeled by the lexicographic order in the range $\{0,|G|-1\}$ on the x-axis. We also present the average fidelity $\bar{\mathcal{F}}$ over all group elements in \fig{fidelity}. The fidelity of an X gate is found to be $\sim$97\% on \texttt{Baiwang}, and CNOT gate to be around $\sim$95\%. For the $\btt$ case, different X gates are needed to prepare state, which are denoted with different shapes in the upper plot of \fig{fidelity}. This variance in state preparation causes noticeable variations in $\mathcal{F}$ observed in \fig{fidelity}. 

We found that for $|g\rangle $ in $ \btt $, the error rate grows slightly with number of X gates in $\mathfrak{U}^{\rm trans}_{-1}$. We can achieve $\bar{\mathcal{F}}=40^{+5}_{-4}\%$ for $\btt$.  Previously, the fidelity of a 5-qubit ordered product encoding $\mathfrak{U}_{-1}$ was benchmarked on \texttt{ibm\_nairobi} with higher connectivity and Pauli twirling used for error mitigation~\cite{Gustafson:2022xdt}. Despite these striking differences, a value of $\bar{\mathcal{F}}=37.0^{+8}_{-8}\%$ was found, suggesting that effective digitizations depend on connectivity and noise channels in nontrivial ways, and therefore should not be neglecting in the discussion of digitization.  

For $\bi$, $\mathfrak{U}^{\rm trans}_{-1}$ can require varying numbers of CNOTs depending on the transpilation, including the orders of mapping matrix elements to the qubits. We show the fidelities for each group element of $\bi$ in \fig{fidelity} (bottom) with shapes indicating the number of CNOTs in the transpiled circuits. As more than 30 CNOT gates are involved, noises from CNOT gates contribute mostly to the quantum errors which reduce the average fidelity to only $\bar{\mathcal{F}} = 4\%$. This low fidelity could be improved by improving transpilation that reduces CNOT counts and implementing error mitigation strategies such as Pauli twirling which have proven effective in the past for LGT~\cite{Yeter-Aydeniz:2022vuy,Carena:2022kpg,Farrell:2022wyt,Gustafson:2022xdt,Atas:2022dqm,Charles:2023zbl,Farrell:2024fit}.

\section{Conclusions and outlook}
\label{sec:conclusion}
In this work, we introduced the block encoding method  -- a general method for digitizing discrete groups on quantum computers and developed the primitive gate set for the two important discrete subgroup of $SU(2)$, including the first implementation ever of $\bi$. The realization of quantum circuits for this largest crystal-like subgroup of $SU(2)$ allows simulating physics of $SU(2)$ deep into the scaling regime.  We have shown that the qubit and T costs as well as robustness to noise are comparable to other digitizations of $SU(2)$. In particular both theoretical analysis and experimental results support the idea that the connection between number of qubits/gates and the true performance in a noisy environment is nontrivial, and such metrics should be considered in evaluating digitization schemes. 

A number of directions for research exist following these results.  Primary is that given its predominance in the gate costs, determination of a quantum Fourier transformation gate within the block encoding method would be invaluable, as it could radically improve the resources costs. Further, given the block encoding method separates group registers into registers of finite fields, it may greatly benefit from formulations on qudit-based devices similar to~\cite{Gustafson:2021jtq,Gustafson:2021qbt,Popov:2023xft,kurkcuoglu2022quantum,Calajo:2024qrc,Gonzalez-Cuadra:2022hxt, illa2024qu8its, Zache:2023cfj,Gustafson:2024kym,Roy:2024uro}.  Finally, the outstanding goal of the discrete group approximation is the largest crystal-like subgroup of $SU(3)$ -- $\vg$ -- which similar to $\bi$ faces obstacles in the order product encoding.  Given it is isomorphic to a subgroup of $GL(3,4)$, the block encoding method provides an avenue for encoding it onto nine 9 ququarts or 18 qubits.

\begin{acknowledgments}
This work is supported by the Department of Energy through the Fermilab QuantiSED program in the area of ``Intersections of QIS and Theoretical Particle Physics". Fermilab is operated by Fermi Research Alliance, LLC under contract number DE-AC02-07CH11359 with the United States Department of Energy. Y.-Y. L is supported by the NSF of China through Grant No. 12305107, 12247103. 
J.S. is supported by Peking University under startup Grant No. 7101302974 and the National Natural Science Foundation of China under Grants No. 12025507, No.12150015; and is supported by the Key Research Program of Frontier Science of the Chinese Academy of Sciences (CAS) under Grants No. ZDBS-LY-7003.
This work was supported by the Munich Institute for Astro-, Particle and BioPhysics (MIAPbP), which is funded by the Deutsche Forschungsgemeinschaft (DFG, German Research Foundation) under Germany's Excellence Strategy–EXC-2094–390783311.
\end{acknowledgments}

\bibliography{refs}

\appendix
\begin{widetext}
\section{Primitive Gates for \texorpdfstring{$\btt$}{BT} Group with Two's Component Encoding}\label{app:bt2}
In this section, we present the primitive gates for $\btt$ group with two's component encoding, e.g. $\left|a = 0\right>  =\left|00\right>, \left|a = 1\right>  =\left|01\right>$ and $\left|a = -1\right>  =\left|11\right>$. Given the addition and multiplication table below, the quantum circuit to implement matrix element additions $\mathfrak{U}_{\rm a}$ and productions $\mathfrak{U}_{\rm p}$ are shown in \fig{btap}. The construction of the group multiplication circuit $\mathfrak{U}_\times$ is similar as \fig{btmulti} in the main text. The inverse and trace operations follow the same rules as 
\eq{btmin} and \eq{tr} in the main text, but with different quantum circuits given in \fig{btintr}.

\begin{minipage}{\textwidth}
 \begin{minipage}[t]{0.47\textwidth}
  \centering
     \makeatletter\def\@captype{table}\makeatother\caption{Addition.}
\begin{tabular}{c|ccc}
\hline
& 00 & 01 & 11 \\
\hline
00 & 00 & 01 & 11 \\
01 & 01 & 11 & 00 \\
11 & 11 & 00 & 01 \\
\hline
\end{tabular}
  \end{minipage}
  \begin{minipage}[t]{0.47\textwidth}
   \centering
        \makeatletter\def\@captype{table}\makeatother\caption{Multiplication.}
\begin{tabular}{c|ccc}
\hline
& 00 & 01 & 11 \\
\hline
00 & 00 & 00 & 00 \\
01 & 00 & 01 & 11 \\
11 & 00 & 11 & 10 \\
\hline
\end{tabular}
   \end{minipage}
\end{minipage}

\begin{figure}[ht]
    \centering
\begin{minipage}[h]{0.15\linewidth}
\vspace{0pt}
\includegraphics[width=\linewidth]{add1}
\end{minipage}
= 
\begin{minipage}[h]{0.3\linewidth}
\vspace{0pt}
\includegraphics[width=\linewidth]{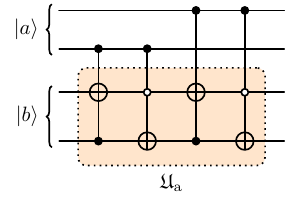}
\end{minipage}
~
\begin{minipage}[h]{0.15\linewidth}
\vspace{0pt}
\includegraphics[width=\linewidth]{prod1}
\end{minipage}
= 
\begin{minipage}[h]{0.3\linewidth}
\vspace{0pt}
\includegraphics[width=\linewidth]{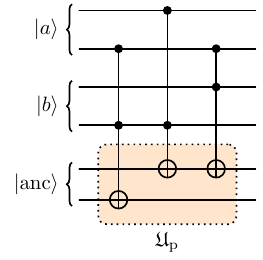}
\end{minipage}

\caption{Quantum circuit implementing matrix element additions and productions in $\mathbf{F}_5$.}
\label{fig:btap}
\end{figure}

\begin{figure}[ht]
  \centering
   \begin{minipage}[h]{0.12\linewidth}
	\vspace{0pt}
	\includegraphics[width=\linewidth]{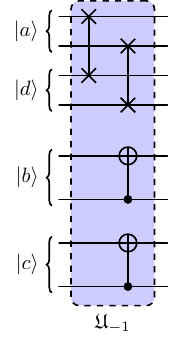}
   \end{minipage}
   ~\begin{minipage}[h]{0.7\linewidth}
	\vspace{0pt}
	\includegraphics[width=\linewidth]{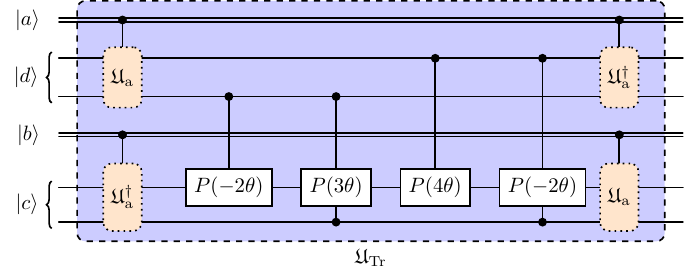}
   \end{minipage}
  \caption{Inverse gate $\mathfrak{U}_{-1}$ and trace gate $\mathfrak{U}_{\rm Tr}$ for $\btt$ group with the two's component encoding.}
    \label{fig:btintr}
\end{figure}

\end{widetext}

\end{document}